\documentstyle[11pt]{article}
\begin{document}

\title{Ordinary differential equations which linearize on differentiation}

\author{{\Large E.V. Ferapontov  and S.R. Svirshchevskii } \\
\ \\
    Department of Mathematical Sciences \\
    Loughborough University \\
    Loughborough, Leicestershire LE11 3TU \\
    United Kingdom \\
    e-mail:
    {\tt E.V.Ferapontov@lboro.ac.uk} \\
    and \\
    Keldysh Institute of Applied Mathematics \\
    Russian Academy of Sciences\\
    Miusskaya Sq. 4, 125047 Moscow \\
    Russia \\
    e-mail:
    {\tt svr@keldysh.ru}
}
\date{}
\maketitle

\newtheorem{theorem}{Theorem}
\newtheorem{proposition}{Proposition}
\newtheorem{lemma}{Lemma}

\pagestyle{plain}

\maketitle

\begin{abstract}

\bigskip

In this short note we discuss ordinary differential equations (ODEs)  which linearize upon one (or more) differentiations. Although the subject is fairly elementary, equations of this type arise naturally in the context of integrable systems.



\end{abstract}

\section{Introduction}

Let us consider a linear $n$-th order ODE  with the general
solution
\begin{equation} u(x)=a_1f_1(x)+...+a_nf_n(x),
\label{u}
\end{equation}
which is a linear superposition of $n$ linearly independent
solutions $f_i(x)$. Imposing a non-linear relation among the
coefficients,
$$
F(a_1,  ..., a_n)=0,
$$
 one obtains an $(n-1)$-parameter family of functions $u(x)$ which automatically solve a
 non-linear ODE of the order $n-1$. By construction,  this  ODE linearizes on differentiation.
 Imposing two relations among the coefficients, one obtains an ODE of the order $n-2$
 which linearizes on two differentiations, etc. Using this simple recipe one can generate
 infinitely many examples of linearizable equations. This note was motivated by the observation that equations of this type arise
 {\em naturally} 
 in the context of integrable systems. The paper is organised as follows.

\noindent Sect. 2 contains a list  of  examples of ODEs which linearize on  differentiation.
These equations appear   in the construction of exact solutions of integrable PDEs, in the classification of integrable hydrodynamic chains, etc.

\noindent In Sect. 3 we derive necessary and sufficient conditions for an ODE to linearize upon a finite number of differentiations.

\noindent In Sect. 4 the general form of non-linear ODEs
linearizable by a differentiation is discussed. It is obtained by
imposing non-linear constraints among  first integrals of a linear
equation. Parallels with the theory of linear invariant subspaces of nonlinear differential operators are briefly discussed.

We point out that the problem of  linearization of nonlinear ODEs
has attracted a lot of attention in the literature. The conditions
of point and contact linearizability of second  and third order
ODEs were first studied by Lie \cite{Lie}, see also \cite{CRCH1}
(p. 38), \cite{CRCH3} (p. 202), \cite{Doubrov, EWLE, Ibragimov}
and references therein. A short remark on `integration via
differentiation' can be found in Kamke \cite{Kam}, sect. 4.14. We
emphasize  that in this paper we are concerned with a different
concept of  `linearizability on differentiation'. Notice that while the
 linearizability by point or contact transformations
is closely connected with symmetry properties of the equation, this
is not true in our case: the symmetry group of the
original nonlinear equation can be trivial, becoming non-trivial for a linear
equation obtained on differentiation.

\section{Examples}

\noindent{\bf Example 1}.
As shown in \cite{Hoppe}, the construction of `follyton' solutions of a nonlinear system associated with  a fourth order self-adjoint spectral problem, reduces to an ODE
$$
u''''u-u'''u'+\frac{1}{2}u''^2=\frac{1}{2}c^4u^2,
$$
$c=const$. On differentiation this equation becomes  linear, $u'''''=c^4u'$, with the general solution
$$
u=a_0+a_1\sinh{cx}+a_2\cosh{cx}+a_3\sin cx+ a_4 \cos cx.
$$
The substitution of this ansatz  into the equation leads to a single quadratic relation among the coefficients,
$
-a_1^2+a_2^2+a_3^2+a_4^2=\frac{1}{2} a_0^2.
$

\medskip

\noindent{\bf Example 2}.  The classification of integrable Hamiltonian hydrodynamic chains associated with the Kupershmidt-Manin  bracket reduces, in a particular case, to a solution of the nonlinear ODE \cite{Fer1}
$$
\begin{array}{c}
(4c^2x^2 u'-12c^2x u -1-\alpha x+2cx^2)u''' +(\alpha-4c x +12c^2u +4c^2x u'-2c^2x^2 u'')u''+ \\
\ \\
(4c-8c^2 u')u' -\frac{1}{2} =0,
\end{array}
$$
here $c, \alpha$ are arbitrary constants. Remarkably, this complicated equation linearizes on  differentiation, taking the form
$$
(4c^2x^2 u'-12c^2x u -1-\alpha x+2cx^2)u''''=0.
$$
Leaving aside the possibility that the coefficient  at $u''''$ equals zero (see \cite{Fer1} for a complete analysis), we
conclude that $u$ must be a cubic polynomial,
$$
u=a_0 +a_1 x + a_2 x^2+a_4 x^3,
$$
where the constants satisfy a single relation $12a_4-8ca_1+16c^2(a_1^2-3a_2 a_0)-4a_2 \alpha+1=0$.

\medskip

\noindent{\bf Example 3}.  Another subclass of integrable hydrodynamic chains from \cite{Fer1} is governed by the ODE
$$
8x^2u'''u'+8xu''u'-4x^2u''^2-u'^2-12u=0
$$
which linearizes on  differentiation,
$$
8x^2u''''+24xu'''+6u''-12=0.
$$
The general solution is given by the formula
$$
u=x^2+a_0+a_1 x+a_2 x^{1/2}+a_3 x^{3/2}
$$
where the constants  satisfy a single quadratic relation
$12a_0+a_1 ^2 -3a_2a_3=0$.

\medskip

\noindent{\bf Example 4}. One of the versions of  equations of associativity \cite{Dubr} reads as
$$
F_{\xi \xi \xi} F_{\eta \eta \eta}-F_{\xi \xi \eta}F_{\xi \eta \eta}=1.
$$
Looking for solutions in the form $F=\xi^3u(x), \ x= \eta/\xi$, one arrives at the ODE
$$
6uu'''-4xu'u'''+2xu''^2-2u'u''=1,
$$
which takes the form $(6u-4xu')u''''=0$ after a  differentiation. The case  $u''''=0$ leads to the general solution
$$
u=a_0+a_1x+a_2x^2+a_3x^3
$$
where the constants satisfy a single quadratic relation $9a_0a_3-a_1a_2=1$. In terms of $F$ these solutions correspond to  polynomials cubic in $\xi$ and $ \eta$.  The case $6u-4xu'=0$ leads to $u=cx^{3/2}$. The corresponding $F$ is given by the formula $F=c(\xi \eta)^{3/2}$ where $c=i\frac{2\sqrt2}{3}$.

\medskip

\noindent{\bf Example 5}. The third order ODE,
\begin{equation}
u'''=s u''\frac{(s+1)u'-2x u''}{(s+1)((s+2)u-2x u')},
\label{s}
\end{equation}
$s=const$, arises in the classification of integrable Hamiltonian hydrodynamic chains associated with Kupershmidt's brackets \cite{Fer2}. It possesses a remarkable property: for  parameter values  $s=1, 2, 3, ...$ this equation linearizes on exactly $s$ differentiations. Thus,
for
$s=1$ the differentiation of (\ref{s}) implies $u''''=0$, so that the general solution is
$$
u=a_0+3a_1x+3a_2x^2+a_3x^3,
$$
where the constants $a_i$ satisfy a single quadratic constraint $a_0a_3-a_1a_2=0$.
For $s=2$, differentiating (\ref{s}) twice, we arrive at $u^{(5)}=0$ with the general solution
$$
u=a_0+4a_1x+6a_2x^2+4a_3x^3+a_4x^4,
$$
where the constants $a_i$ satisfy a system of quadratic constraints
$$
a_0a_3-a_1a_2=0, ~~~ a_1a_4-a_2a_3=0, ~~~ a_0a_4-a_2^2=0;
$$
notice that these constraints specify a determinantal variety characterized by the requirement that the rank of the matrix
$$
\left(
\begin{array}{ccc}
a_0 & a_1 & a_2 \\
a_2& a_3 & a_4
\end{array}
\right)
$$
equals one. The mystery of this example is unveiled by the  formula for its general solution,
$$
u=a(x+c)^{s+2}+b(x-c)^{s+2},
$$
which is valid for any $s$; here $a, b, c$ are arbitrary constants
(we thank A.P. Veselov for this  observation).

\section{Necessary and sufficient conditions for  the linearizability}

In this section we demonstrate how to derive necessary and sufficient conditions for a non-linear ODE to linearize on one (or more) differentiations. The procedure is fairly straightforward and  can be readily adapted to particular situations.

\subsection{First order ODEs which linearize on one differentiation}

Let us characterize first order equations
$$
u'=f(x, u)
$$
which imply a linear equation,
$$
u''=a(x)u'+b(x)u+c(x),
$$
on one differentiation. Thus, we have $f_x+f_uf=af+bu+c$. Differentiating
this relation twice with respect to $u$, and  introducing $F=f_x+f_uf$, one obtains $F_{uu}=af_{uu}$. Differentiating this by $u$ once again one has $F_{uuu}=af_{uuu}$. Thus, the required linearizability condition takes the form
$$
F_{uuu}f_{uu}=f_{uuu}F_{uu};
$$
see Sect. 4 for the  general (implicit) form of all such right hand sides $f(x, u)$ (formula (\ref{F12})).

\subsection{Second order ODEs which linearize on one differentiation}

Let us characterize second order equations
$$
u''=f(x, u, p), ~~~ p=u',
$$
which imply a linear equation,
$$
u'''=a(x)u''+b(x)u'+c(x)u+k(x),
$$
on one differentiation. Thus, we have $f_x+f_up+f_pf=af+bp+cu+k$.  Applying to this relation the operators $\partial_u^2, \ \partial_u\partial_p, \ \partial_p^2$, and  introducing $F=f_x+f_up+f_pf$, one obtains
$$
F_{uu}=af_{uu}, ~~~ F_{up}=af_{up}, ~~~F_{pp}=af_{pp},
$$
or, equivalently, $d^2F=a d^2f$ (here the second symmetric differential $d^2$ is calculated with respect to $u$ and $p$ only). Differentiating this once again by $u$ and $p$ one  obtains $d^3F=a d^3f$. Thus, the required linearizability condition takes the form
$$
d^3F\ d^2f=d^3f\ d^2F.
$$

\subsection{First order ODEs which linearize on two differentiations}

Here we characterize first order equations
$$
u'=f(x, u)
$$
which imply a linear equation
$$
u'''=a(x)u''+b(x)u'+c(x)u+k(x)
$$
after two differentiations. Introducing $F=f_x+f_uf$ and $G=F_x+F_uf$, we have
$G=aF+bf+cu+k$. Differentiating this twice with respect to $u$ one obtains $G_{uu}=aF_{uu}+bf_{uu}.$
This implies $G_{uuu}=aF_{uuu}+bf_{uuu}$ and $G_{uuuu}=aF_{uuuu}+bf_{uuuu}$. Thus, the required condition is
$$
\mbox{det} \left[ \begin{array}{ccc}
 G_{uu} & F_{uu} & f_{uu} \\
  G_{uuu} & F_{uuu} & f_{uuu} \\
 G_{uuuu} & F_{uuuu} & f_{uuuu}
 \end{array}
 \right]=0.
$$
In all of the above examples, the linearizability is characterized by differential relations which must be satisfied by the right hand side of the equation. As we demonstrate in the next section, these differential equations can be integrated in closed form, leading to (implicit) representation for all linearizable equations.

\section{General form of  linearizable equations}

All equations linearizable by a differentiation can be obtained by imposing functional
relations among first integrals of  linear equations. Since the
first integrals can be parametrized explicitly by arbitrary
functions of the independent variable $x$, this provides a general
formula for equations which linearize on  differentiation.

\noindent {\bf Example 6.} Let us describe all ODEs which reduce to
$u''''=0$ after one or two differentiations. The basis of  first
integrals consists of
\begin{equation}
I_1=u''', ~~~ I_2=xu'''-u'', ~~~ I_3=x^2u'''-2xu''+2u', ~~~
I_4=x^3u'''- 3x^2u''+6xu'-6u. \label{I4}
\end{equation}
Any third order equation which reduces to $u''''=0$ after one
differentiation can be represented by a single relation among the
first integrals,
$$
F(I_1, I_2, I_3, I_4)=0.
$$
Any second order equation which reduces to $u''''=0$ after two
differentiations can be represented in implicit form by two
relations,
$$
F(I_1, I_2, I_3, I_4)=0, ~~~ G(I_1, I_2, I_3, I_4)=0;
$$
one has to eliminate $u'''$ to obtain the  required second order
equation.
\medskip

In general, let us consider a linear ODE
\begin{equation}
L[u] \equiv u^{(n)}+b_{1}(x)u^{(n-1)}+\dots+b_{n-1}(x)u^{\prime}+
b_{n}(x)u=b(x) \label{L}.
\end{equation}
Let $f_0(x)$ be its particular solution, and let $f_1(x),\dots,
f_n(x)$ be a fundamental system of solutions (FSS) of the
corresponding homogeneous equation. A complete set of first
integrals for the equation (\ref{L}) can be taken in the form
\begin{equation} I_{i}[u]=\frac {W[f_{1},\,
...\,,f_{i-1},u-f_0,f_{i+1},\, ... \,,f_{n}] }{ W[f_{1},\, ...
\,,f_{n}]},\quad i=1,\, ... \,,n, \label{FI}
\end{equation}
where $W[ \cdot \,]$ denotes the Wronskian of the functions
indicated in square brackets. Indeed, for an arbitrary solution
$f(x)=f_0(x)+a_1 f_1(x)+ ... + a_n f_n(x)$ we have $I_i [f(x)] =
a_i$ for $i=1,\, ... \,,n$, i.e., all $\{I_i\}$ take constant values.
Applying (\ref{FI}) to the equation $u''''=0$ with $f_0(x)=0$,
$f_i(x)=x^{i-1}$, $i=1,...,4$, one obtains  first integrals which
coincide with (\ref{I4}) up to constant factors.

For example, in the case $n=2$ the general form of
first order equations linearizable by one differentiation is
represented via arbitrary  functions $f_0(x)$, $f_1(x)$,
$f_2(x)$ and $F$ as
\begin{equation}
F(I_1[u], I_2[u])=0
 \label{F12}
\end{equation}
where
$$
  \matrix{
 I_{1}[u]=\frac {W[u-f_0,f_{2}] }{
W[f_{1},f_{2}]}=\frac {(u-f_0) f'_{2} - (u-f_0)' f_{2} }{f_1 f'_2
- f'_1 f_{2}}, \quad  
 I_{2}[u]=\frac{W[f_{1},u-f_0] }{ W[f_{1},f_{2}]}=\frac{ - (u-f_0) f'_1 + (u-f_0)' f_1 }{f_1 f'_2 - f'_1
 f_{2}}.
 }
$$
The differentiation of  (\ref{F12}) yields
$$
\frac {- f_2 F_{I_{1}} + f_1 F_{I_{2}}}{(W[f_{1},f_{2}])^2} \Bigl(
(u-f_0)''W[f_{1},f_{2}] - (u-f_0)'(W[f_1,f_2])' + (u-f_0)
W[f'_1,f'_2] \Bigr)=0,
$$
leading to the linear equation
$$
W[f_1,f_2,u-f_0]=0, \quad \mbox{or} \quad  W[f_1,f_2,u] =
W[f_1,f_2,f_0].
$$
This construction  generalizes to the case of arbitrary $n$ in a straightforward way.
In particular,  equation (\ref{F12}) provides the general (implicit) form for 
equations discussed in the Subsection 3.1. Similar representations
can be obtained for all other cases from Section 3.

The linear span of the functions $f_{i}(x)$,
\begin{equation}
W_{n}={\it L}\{f_{1}(x),\dots,f_{n}(x)\},
\label{W}
\end{equation}
 represents the linear space of solutions to the homogeneous equation
 $L[u]=0$ corresponding to (\ref{L}).
The space $W_{n}$ is said to be invariant with respect to a
differential operator $F$, if $F[W_{n}] \subseteq W_{n}$. A
systematic study of operators preserving a given subspace was
initiated in \cite{Gal} in the context of constructing explicit
solutions for nonlinear evolution equations. The general form of
operators preserving the subspace (\ref{W}) is given by
\begin{equation}
F[u]=\sum_{i=1}^{n} A^{i}(I_{1},\dots,I_{n})f_{i}(x), \label{F}
\end{equation}
where $A^{i}(I_{1},\dots,I_{n})$ are arbitrary functions of the
first integrals of the equation $L[u]=0$ 
(see \cite{Svr} and \cite{GalSvr} for more details). Given an
operator $F$ of the form (\ref{F}), we introduce the equation
\begin{equation}
F[u]=0,
\label{F0}
\end{equation}
 and look for its solutions in the form (\ref{u}). The substitution of
 (\ref{u}) into (\ref{F0}) yields the identity
$$
\sum_{i=1}^{n} A^{i}(I_{1},\dots,I_{n})f_{i}(x)=0
$$
implying that
\begin{equation}
A^{i}(I_{1},\dots,I_{n})=0  ~~ \mbox{for} ~~ i=1,...,n. \label{A}
\end{equation}
This system imposes  relations on the coefficients $a_i$.

Most of the previous examples fit into this scheme. For instance,
the equations from examples 2 and 4 are written via the first
integrals (\ref{I4}) as
$$
F[u] \equiv 2C^2(I_4I_2-I_3^2)+2CI_3-I_1-\alpha I_2-\frac{1}{2}=0
$$
and
$$
F[u] \equiv -I_4 I_1+I_3 I_2 - 1 = 0,
$$
respectively, where $I_i$ are given by (\ref{I4}). In both cases
the system (\ref{A}) reduces to a single relation.

Similarly, for the equation (\ref{s}) rewritten as
\begin{equation}
F_s[u] \equiv (s+1)[(s+2)u'''u-su''u']-2x[(s+1)u'''u'-su''^2]=0,
\label{Fs}
\end{equation}
one obtains, for $s=1$,
$$
F_1[u] \equiv   \frac{1}{2} (  I_2I_3 - I_1I_4 ) = 0;
$$
this again leads to a single relation  (\ref{A}).
In the case $s=2$ we have a representation
\begin{equation}
F_2[u] \equiv  \frac{1}{2} [ (I_2I_3 -I_1I_4) x^2 + (I_1I_5
-I_3^2) x + (I_3I_4 -I_2I_5) ] = 0 \label{s2}
\end{equation}
via the first integrals $I_1,...,I_5$ of the equation $u^{(5)}=0$. In
this case the system (\ref{A}) formally consists of three relations,
$$
I_2I_3 -I_1I_4 = 0, ~~~ I_1I_5 -I_3^2 = 0, ~~~ I_3I_4 -I_2I_5 = 0,
$$
however,  only two of them are functionally independent.


\noindent {\bf Remark.} In accordance with \cite{Svr2}, every
operator (\ref{F}) of the order $n-1-k$, admitted by the equation
\begin{equation}
u^{(n)}=0,
\label{un}
\end{equation}
is expressed in terms of the differences
\begin{equation}
J^k_i = x J^{k-1}_{i}-J^{k-1}_{i+1}, \quad i=1,\dots,n-k,
\label{J}
\end{equation}
with $J_i^0 \equiv I_i$, $i=1,\dots,n$ ( $I_i$ are first integrals
for (\ref{un})). All these expressions are of the order $n-1-k$,
and satisfy the identity $D^{k+1} J^k_i = 0$ on solutions of the
equation (\ref{un}). Setting $n=s+3$, $k=s-1$, we obtain that any
third-order operator admitted by the equation $u^{(s+3)}=0$ is
defined via the functions $J^{s-1}_i$, $i=1,\dots,4$. For
instance, the operator $F_s$ from (\ref{Fs}) is represented as
$$
F_s[u] = \frac {1}{2} (J_2^{s-1}J_3^{s-1} - J_1^{s-1}J_4^{s-1}).
$$

\section{Acknowledgements}
We thank A Aksenov, V Galaktionov, J Hoppe, N Ibragimov and A Veselov for clarifying discussions and interest. The research of  SRS  was supported by the RFBR grant 06-01-00707.

\end{document}